\documentclass[11pt,a4paper,epsfig]{article}

\usepackage{amsfonts}
\usepackage{graphicx}
\usepackage[colorlinks=true,linkcolor=blue]{hyperref}

\title{\textbf{Kink fluctuation asymptotics and zero modes}}

\author{A. Alonso Izquierdo$^{(a)}$, and J. Mateos Guilarte$^{(b)}$
\\ {\normalsize {\it $^{(a)}$ Departamento de Matematica
Aplicada and IUFFyM}, {\it Universidad de Salamanca, SPAIN}} \\ {\normalsize {\it $^{(a)}$ Departamento de Fisica
Fundamental and IUFFyM}, {\it Universidad de Salamanca, SPAIN}}}

\date{}

\begin{document}

\maketitle

\begin{abstract}
In this paper we propose a refinement of the heat kernel/zeta function treatment of kink quantum fluctuations in scalar field theory, further analyzing the existence and implications of a zero energy fluctuation mode.  Improved understanding of the interplay between zero modes and the kink heat kernel expansion delivers asymptotic estimations of one-loop kink mass shifts with remarkably higher precision than previously obtained by means of the standard
Gilkey-DeWitt heat kernel expansion.
\end{abstract}


\section{Introduction}

 In 1961 Skyrme \cite{Skyrme} proposed the following bold idea about the bridge
between Classical and Quantum Field Theory: \lq\lq If plane wave solutions of the field equations in the classical domain emerge as light mesons in the quantum domain, solitary wave solutions make the quantum leap, becoming heavy baryons in a non-linear quantum field theory".   This image from hadronic physics prompted the task of studying the quantum nature and properties of topological and non-topological classical non-linear waves with far-reaching consequences, ranging from Cosmology and Gauge Theories to Condensed Matter Physics. In 1962, see \cite{PS}, Skyrme, together with Perring, found that the sine-Gordon equation with its roots in the Geometry of Surfaces of constant (negative) curvature offered an ideal playground in which Skyrme's idea was materialized analytically.
In the mid seventies, Dashen, Hasslacher and Neveu addressed the investigation of one-loop kink fluctuations by developing the $\hbar$-expansion of the (1 + 1)-dimensional $\phi^4$ and sine-Gordon scalar field models. In the two seminal papers \cite{Dashen1974,Dashen1975}, without and with fermionic fields, the authors encoded their findings in the evaluation of kink mass quantum shifts up to one-loop order, as well as the computation of the semi-classical/WKB corrections to the Bohr-Sommerfeld spectra of the sine-Gordon breathers.  In the review reports \cite{Alonso2006}  and \cite{Guilarte2009} later developments on the subject achieved by several research groups in the USA and the EU  between 1998 and 2006 are reported and summarized. This renaissance in the analysis of topological soliton fluctuations addressed mainly, but not only, supersymmetric versions of the same type of phenomena and/or generalizations to models with several scalar fields, also touching  on the very delicate point of the influence of boundary conditions and  their interplay with supersymmetry plus the choice of renormalization criteria.

Here  we shall  concentrate on an specific method to deal with kink fluctuations. The one-loop kink mass shift induced by the kink quantum fluctuations is due to three contributions: 1) The kink zero point energy collecting the energy of the kink ground state where all the fluctuation modes are unoccupied. 2) The analogous vacuum zero point energy that must be subtracted. 3)  The energy induced by the one-loop
renormalization mass counter-term on the kink background (measured with respect to the same effect on the vacuum). All these three terms are divergent and it was decided in \cite{Bordag2002} -supersymmetric kinks- and \cite{Alonso2002} -bosonic kinks- to regularize these quantities by means of the spectral zeta function of the kink and vacuum second-order fluctuation operators. In the sine-Gordon and $\phi^4$ models the kink Hessian operators are Schrodinger operators of the P$\ddot{\rm o}$schl-Teller type whose spectral problem is solvable. Therefore, one can exactly compute the kink spectral zeta function.

In other interesting models the spectral information about the differential operators governing kink fluctuations is usually grossly insufficient for gaining analytical knowledge of the spectral zeta functions. The idea in the references quoted above was to rely on the determination of the spectral zeta function via the Mellin transform of the heat trace of the same operator and then use the asymptotic expansion of this latter spectral function to obtain information about the kink fluctuation asymptotics.  With different stimuli from Mathematics \cite{Gilkey1984, Roe1988} and Physics  \cite{DeWitt1965}, interesting developments concerning the high-temperature expansion of the kernel of the generalized heat equation associated with a differential operator of the Laplace or Dirac type took appeared in the mid sixties. In particular, Gilkey, following classical works on the heat kernel proof of the index theorem, unveiled the meromorphic structure of the generalized zeta function, showing that the residua at the poles of the spectral zeta function, determined from the Seeley coefficients of the heat kernel expansion, are certain topological invariants/characteristic classes. In turn DeWitt used the heat kernel expansion to deal with quantum fields on curved backgrounds.  During the last twenty-five years the use of heat kernel/zeta function methods in Quantum Field Theory has become a very important area of research: we explicitly mention the reports and textbooks \cite{Avramidi, Elizalde1994, Kirsten2002, Vassilevich2003} for the reader to become acquainted with this important subject. During the last decade we (and our colleagues) profited from the theoretical machinery available by applying these tools at our disposal to express the one-loop kink mass shift as a truncated series in the Seeley coefficients of the kink-Hessian-heat function, see \cite{Alonso2002,Alonso2002A,Alonso2004,Alonso2011, Alonso2012}, in several $(1+1)$-dimensional field theoretical models of a single real scalar field .

Our goal in this paper is to address a very delicate issue. The Gilkey-DeWitt heat kernel expansion  is an \lq\lq inverse temperature" power series such that the zeroth order term is the heat kernel of the \lq\lq free " Laplace/Helmotz operator. Thus, the GDW heat kernel expansion coincides at infinite temperature with the free heat kernel. Nevertheless, because of the completeness of the eigenfunctions of the kink-Hessian
Sturm-Liouville problem the corresponding heat kernel cannot go to zero at low temperature if there are zero modes in the spectrum, as happens in the GDW expansion.  Given that there is always a zero energy kink fluctuation, the translational mode,  here  we propose to modify the GDW heat kernel expansion to incorporate this notion when heat trace/zeta functions techniques are to be applied to the study of kink fluctuations. Even though
the standard lore is that zero modes enter at two-loop order in the $\hbar$-expansion of quantum mass shifts, they play a hidden r$\hat{\rm o}$le noticed, for instance, in the Cahill-Comtet-Glauber computation \cite{Glauber} in terms of the bound state eigen-values and the threshold of the continuous spectrum. Thus, we shall assign special status to the zero modes that, somehow, is tantamount to performing the heat kernel expansion around the sine-Gordon kink Hessian. The new procedure is not  only conceptually more satisfactory but enhances the numerical precision in the computation of the kink mass quantum correction to a remarkable extent.

The organization of this paper is as follows: in Section \S.2 we briefly review the semi-classical kink mass DHN formula and variations of this expression obtained in the heat kernel/zeta function regularization procedure. Section
\S.3 contains the main novelty in this paper: the modified heat kernel expansion adapted to the existence of the kink zero mode is introduced. In the Section \S.4, the new formula is applied to the sine-Gordon and $\phi^4$ kinks to test the modified procedure. The response found in these models is in complete agreement with the exact results, improving the approximation reached in the traditional heat kernel expansion approach. We shall also work this method to evaluate the one-loop mass shifts in the scalar field models recently addressed in \cite{Alonso2012}. The interest in these models lies in the fact that full information about the kink Hessian spectrum allows us to know exactly the quantum correction offering a very good playground to gauge all the methods. Finally, we shall run the new procedure in two one- parametric families of scalar field models recently analyzed in \cite{Alonso2011}: a generalized family of $\phi^6$ models and the double sine-Gordon model. In both cases the DHN formula is not applicable and the exact result is not known. Nevertheless, interesting results can be obtained in the heat kernel approach, old in \cite{Alonso2011} and new in this paper. In the Appendix, a Mathematica code that automatizes the computations involved in the new procedure has been included.

\section{One-loop quantum kink fluctuations }

In this Section we briefly review the standard lore about the conceptual understanding of quantum kink fluctuations as well as the approach of our group to this subject in the heat kernel/zeta function framework. We shall follow the notation and conventions fixed in the recent reference \cite{Alonso2011}.

\subsection{Classical field theoretical models and kinks}

The action governing the dynamics in a $(1+1)$-dimensional relativistic one-scalar field theoretical model is of the form:
\[
\tilde{S}[\psi]=\int \!\! \int\, dy^0dy^1 \, \left(\frac{1}{2}\frac{\partial\psi}{\partial y_\mu}\cdot \frac{\partial\psi}{\partial y^\mu}- \tilde{U}[\psi(y^\mu)] \right) \quad .
\]
Here, $\psi(y^\mu): \mathbb{R}^{1,1} \rightarrow \mathbb{R}$ is a real scalar field; i.e., a continuous map from the $(1+1)$-dimensional Minkowski space-time to the field of the real numbers.  $y^0=\tau $ and $y^1=y$ are local coordinates in ${\mathbb R}^{1,1}$, which is equipped with a metric tensor that we choose: $g_{\mu\nu}={\rm diag}(1,-1)$, $\mu,\nu=0,1$.

We shall work in a system of units where the speed of light is set to one, $c=1$, but we shall keep the Planck constant $\hbar$ explicit because we shall search for one-loop corrections, proportional to $\hbar$, to the classical kink masses.
In this system, the physical dimensions of fields and parameters are:
\[
[\hbar]=[\tilde{S}]=M L \quad , \quad [y_\mu]=L \quad , \quad [\psi]=M^\frac{1}{2}L^\frac{1}{2} \quad , \quad [\tilde{U}]=ML^{-1} \quad .
\]
The models that we shall consider are distinguished by different choices of the part of the potential energy density that is independent of the field spatial derivatives: $\tilde{U}[\psi(y^\mu)]$. In all of them, there will be two special parameters, $m_d$ and $\gamma_d$, to be determined in each case, carrying the physical dimensions: $[m_d]=L^{-1}$ and $[\gamma_d]=M^{-\frac{1}{2}}L^{-\frac{1}{2}}$. We define the non-dimensional coordinates, fields and potential in terms of these parameters:
\[
x_\mu=m_d y_\mu \, \, \, , \, \, \, x_0=t \, \, , \, \, x_1=x \quad , \quad \phi= \gamma_d \psi \qquad , \qquad U(\phi)=\frac{\gamma_d^2}{m_d^2} \tilde{U}(\psi) \qquad .
\]
The action and the \lq\lq static " part of the energy are also proportional to dimensionless action and energy functionals, namely:
\begin{eqnarray}
\tilde{S}[\psi]&=&\frac{1}{\gamma_d^2}S[\phi]= \frac{1}{\gamma_d^2}\int\!\!\!\int\, dx^0dx^1 \, \left[\frac{1}{2}\frac{\partial\phi}{\partial x_\mu}\cdot \frac{\partial\phi}{\partial x^\mu}- U[\phi(x^\mu)] \right] \, \, \, , \label{action}\\\tilde{E}[\psi]&=&\frac{m_d}{\gamma_d^2}E[\phi]= \frac{m_d}{\gamma_d^2}\int\!\! dx \, \left[\frac{1}{2}\left(\frac{d\phi}{d x}\right)^2 +U[\phi(x)] \right] \, \, \, , \, \, \, x^1=x \, \, \, ,\label{energy}
\end{eqnarray}
where we shall assume that $U(\phi)$ is a non-negative twice-differentiable function of $\phi$: $U(\phi)\in C^2({\mathbb{R}})$ and $U(\phi)\geq 0$ for $\phi \in \mathbb{R}$. The configuration space ${\cal C}$ of the system is the set of field configurations ${\cal C}=\{\phi(t_0,x)\in {\rm Maps}(\mathbb{R}^{1},\mathbb{R})/ E[\phi]<+\infty\}$. We assume that the set ${\cal M}=\{\phi^{(i)}\,\,/ \,\,U(\phi^{(i)})=0\}$ of the minima of $U$ is a discrete set. The simplest solutions of the field equations
\begin{equation}
\left(\frac{\partial^2}{\partial t^2}-\frac{\partial^2}{\partial x^2}\right)\phi(t,x)=-\frac{\delta U}{\delta\phi}(t,x) \label{socfe}
\end{equation}
are static and homogeneous, and hence the elements of ${\cal M}$. The small (quadratic) fluctuations around any of these constant solutions, $\phi(t,x)=\phi^{(i)}+\delta\phi(t,x)$, satisfy the linearized field equations:
\begin{equation}
\left(\frac{\partial^2}{\partial t^2}-\frac{\partial^2}{\partial x^2}+v^2\right)\delta\phi(t,x)+{\cal O}[(\delta\phi)^2]=0 \quad \mbox{with} \quad \left.\frac{\partial^2 U}{\partial\phi^2}\right|_{\phi^{(i)}}=v^2 \quad . \label{lpde}
\end{equation}
The solutions of (\ref{lpde}), of the form $\delta\phi_k(t,x)=e^{i\nu(k)t}f_k(x)$, are the normal modes of fluctuation of the system near one minimum of $U$. These linear waves are built from the eigenfunctions of the second-order \lq\lq vacuum"{\footnote{ We pass to use QFT terminology: the linear waves in the quantized theory are the light mesons, whereas the minima of $U$ are the vacua of the system. }} fluctuation differential operator:
\begin{equation}
K_0=-\frac{d^2}{dx^2}+v^2 \, \, \, , \, \, \, K_0f_k(x)=\nu^2(k)f_k(x) \, \, \, , \, \, \, f_k(x)=e^{ikx}\, \, , \,
k\in{\mathbb R} \, \, \, , \, \, \, \nu^2(k)=k^2+v^2 \quad .
\label{operatorK0}
\end{equation}
 We remark on a point usually unnoticed: there is a \lq\lq half-bound" state at the threshold of the continuous spectrum $v^2$ with constant eigenfunction.

If the cardinal of ${\cal M}$ is greater than 1, there may exist spatially dependent static solutions of the solitary wave type (non-dispersive non-linear waves), which in the literature are referred to as kinks or lumps, see e.g. \cite{Rajaraman1982, Drazin1996, Manton2004}. The static kink solutions $\phi_K(x)$ satisfy the first-order ODE
{\footnote{The kinks are also static solutions of the PDE equation (\ref{socfe}).}}
\begin{equation}
\frac{d\phi}{dx} = \pm\sqrt{2 U(\phi)}
\label{ode1} \quad ,
\end{equation}
together with the asymptotic conditions guaranteeing finiteness of the energy:
\begin{equation}
\lim_{x\rightarrow  +\infty} \phi_K(t,x)=\phi^{(i)}\in {\cal M}\hspace{0.5cm},\hspace{0.5cm}  \lim_{x\rightarrow  -\infty} \phi_K(t,x)=\phi^{(i\mp 1)}\in {\cal M} \hspace{0.5cm},\hspace{0.5cm} \lim_{x\rightarrow \pm \infty} \frac{\partial \phi_K(t,x)}{\partial x}=0 \, \, .
\label{asymptotic}
\end{equation}
 Small fluctuations around any of the equivalent kink/antikink solutions, $\phi(t,x)=\phi_K(x)+\delta\phi(t,x)$, still solving (\ref{socfe}) satisfy the linearized field equations:
\begin{equation}
\left(\frac{\partial^2}{\partial t^2}-\frac{\partial^2}{\partial x^2}+v^2+V(x)\right)\delta\phi(t,x)+{\cal O}[(\delta\phi)^2]=0 \quad , \quad \left.\frac{\partial^2 U}{\partial\phi^2}\right|_{\phi_K(x)}=v^2+V(x) \quad . \label{lpdek}
\end{equation}
Again the solutions of (\ref{lpdek}), of the form $\delta\phi_k(t,x)=e^{i\omega(q)t}f_q(x)$, are the normal modes of kink fluctuations coming from the eigenfunctions of the second order kink fluctuation differential operator:
\begin{equation}
K=-\frac{d^2}{dx^2}+v^2+V(x) \,\,\, .
\label{operatorK}
\end{equation}
We shall assume that the potential well goes to the same limit at both ends of the straight line: $\lim_{x\rightarrow \pm \infty} V(x) =0$.
This is true in all models for which $\frac{\partial^2 U}{\partial\phi^2} |_{\phi^{(i)}}= v^2= \frac{\partial^2 U}{\partial\phi^2}|_{\phi^{(i+1)}}$, where $\phi^{(i)}$ and $\phi^{(i+1)}$ are the vacuum solutions that are connected by the kink. The kink normal modes built from the eigenfunctions of $K$ as a linear superposition
\[
Kf_q(x)=\omega^2(q)f_q(x) \quad , \quad \omega^2(q)=q^2+v^2
\]
are not plane waves but some dispersive wave functions distorted by the kink. Contrary to $K_0$, which is
a Laplace/Helmotz operator, $K$ is a Schr$\ddot{\rm o}$dinger operator for a potential well. Therefore, the spectrum of $K$ is formed by scattering states, bound states and, possibly, half-bound states.

The lowest $K$-bound state fluctuation  (mesons trapped by the kink) is always a zero mode: $f_0(x)=\frac{d\phi_K}{dx}$, $\omega_0=0$. This $K$-eigenfunction is no more than the Goldstone boson due to the spontaneous symmetry breaking by the kink of the invariance of the system with respect to spatial translations: $x\to x+a$.

\subsection{Kink mass quantum correction induced by one-loop fluctuations}

\subsubsection{The \lq\lq first" DHN formula}

The problem of computing the shift in the kink mass induced by kink fluctuations in the one-loop order
of the sine-Gordon and $\lambda(\phi^4)_{1+1}$ models was solved by Dashen, Hasslacher and Neveu in \cite{Dashen1974}.
They succeeded in building a formula that collects all the effects contributing to this quantum effect.{\footnote{We
refer to this formula as the first DHN formula, to be distinguished from the \lq\lq second" DHN formula applicable to the quantum sine-Gordon breathers.}} We describe the formula as a regularization-renormalization construction in two steps.

\begin{enumerate}

\item Zero-point kink energy renormalization: mode-by-mode subtraction of the zero point vacuum energy

In the first step, a zero point kink energy  (the energy due to all the $K$-fluctuation modes being unoccupied)  renormalization is performed by means of a mode-by-mode subtraction of the zero point vacuum energy -all the $K_0$-fluctuation modes being unoccupied-. In the Appendix of \cite{Alonso2004}, we generalized the DHN formula to kinks other than the sine-Gordon and $\lambda(\phi^4)_{1+1}$ kinks, using the one-dimensional Levinson theorem, finding the formula:
\begin{equation}
\bigtriangleup E_1(\phi_K)=\frac{\hbar\gamma_d^2}{2}\Big[\lim_{\Lambda\rightarrow \infty}\int_{0}^\Lambda dk \, \frac{1}{\pi}\, \frac{\partial\delta(k)}{\partial k}\sqrt{k^2+v^2}+\frac{1}{2\pi}\langle V(x) \rangle +\sum_{j=2}^{b-1}\, \omega_j+s_b\omega_b-\frac{v}{2}\Big] \label{kcas} \quad ,
\end{equation}
where $\langle V(x) \rangle =\int_{-\infty}^\infty \, dx \, V(x)$ and $\delta(k)$ are the phase shifts associated with the second order kink fluctuation operator (\ref{operatorK}).
The two last terms in (\ref{kcas}) collect the contribution of the possible half-bound state of $K$ and the ever-present half-bound state of $K_0$ and represent the novelty in \cite{Alonso2004} with respect to the DHN formula.

\item Mass renormalization  counter-term contribution:

$\bigtriangleup E_1(\phi_K)$ in (\ref{kcas}) is still divergent because in models with interactions there are more ultraviolet divergences than the vacuum energy. In $(1+1)$-dimensional scalar field theory normal ordering tames all the uv divergences. The ordering induces a self-energy counter-term that, in turn, via the expectation value of the scalar field operator at the kink and vacuum coherent states, produces the following contribution to the one-loop kink mass shift expression, see e.g. \cite{Guilarte2009}:
\begin{equation}
\bigtriangleup E_2(\phi_K) = -\frac{\hbar\gamma_d^2}{8\pi}  \left<V(x) \right>  \lim_{\Lambda\rightarrow \infty}\int_{-\Lambda}^\Lambda \frac{dk}{\sqrt{k^2+v^2}} \quad , \label{kcas2}
\end{equation}
\end{enumerate}

\noindent The addition of these two quantities, previously regularized separately by choosing a cutoff that counts the same number of eigen-modes over the kink as over the vacuum, provides the finite value of the one-loop kink mass shift:
\[
\bigtriangleup E(\phi_K)  = \bigtriangleup E_1(\phi_K) + \bigtriangleup E_2(\phi_K)\quad .
\]

\subsubsection{The zeta function regularization procedure}

In order to apply the above DHN scheme effectively in the computation of one-loop kink masses, it is necessary to know the eigenvalues of the bound states and the scattering wave phase shifts of the operator $K$. Only for the sine-Gordon and $\lambda(\phi)^4_{1+1}$ kink is this information fully available and other approaches to the problem are necessary in the investigation of quantum fluctuations of more complex kinks. The application of the zeta function regularization procedure has proved to be very effective in such a situation.
The vacuum energy induced by quantum fluctuations is first regularized by assigning it the value of the spectral zeta function of the $K_0$-operator (a meromorphic function) at a regular point in $s\in{\mathbb C}$:
\begin{equation}
\bigtriangleup E(\phi^{(i)})=\frac{\hbar \gamma_d^2}{2}\zeta_{K_0}(-{\textstyle\frac{1}{2}}) \, \, \rightarrow \, \, \bigtriangleup E(\phi^{(i)})[s]=\frac{\hbar \gamma_d^2}{2}\frac{\mu}{m_d}\left(\frac{\mu^2}{m_d^2}\right)^s\zeta_{K_0}(s) \quad ,
\end{equation}
where $\mu$ is a parameter of dimensions $L^{-1}$ introduced to keep the dimensions of the regularized energy right. We stress that $s=-\frac{1}{2}$ is a pole of this function. The same rule is applied to control the kink energy divergences by means of the spectral zeta function of $K$ and, finally, the kink Casimir energy is regularized
in the form:
\begin{equation}
\bigtriangleup E_1(\phi_K)[s]= \bigtriangleup E_0(\phi_K)[s]-\bigtriangleup E_0(\phi^{(i)})[s]=\frac{\hbar \gamma_d^2}{2}\left(\frac{\mu^2}{m_d^2}\right)^{s+\frac{1}{2}}\left(\zeta_{K}(s)-\zeta_{K_0}(s)\right) \,\,\, . \label{zrkc}
\end{equation}

The advantage of this procedure is that there is no need for detailed information about the spectrum of $K$ to calculate the spectral zeta function. It is possible to elude this problem in two steps:
\begin{enumerate}
\item First, an asymptotic formula for the $K$-heat trace can be obtained from the GDW heat-kernel expansion in terms of the so-called Seeley coefficients.
\item The Mellin transform of the heat trace in turn provides the $K$-zeta function such that the regularized kink energy reads:
\begin{equation}
\bigtriangleup E_1(\phi_K)[s]=\frac{\hbar \gamma_d^2}{2} \left(\frac{\mu^2}{m_d^2}\right)^{s+\frac{1}{2}}\frac{1}{\Gamma(s)}\left\{\int_0^\infty \, d\beta \, \beta^{s-1}\left(h_{K}(\beta)-h_{K_0}(\beta)\right)\right\} \,\,\, .
\label{zrkc2}
\end{equation}
\end{enumerate}

The energy due to the one-loop mass counter-term (\ref{kcas2}) can be also regularized by the zeta function procedure:
\begin{equation}
\bigtriangleup E_2(\phi_K)[s]=  \frac{\hbar \gamma_d^2}{2}\langle V(x) \rangle \left(\frac{\mu^2}{m_d^2}\right)^{s+\frac{1}{2}}\lim_{l\to\infty}\frac{1}{l}\frac{\Gamma(s+1)}{\Gamma(s)}\zeta_{K_0}(s+1)
 \,\,\, .\label{zrkc3}
\end{equation}
Finally, we write the zeta function regularized DHN formula:
\begin{equation}
\bigtriangleup E(\phi_K)= \lim_{s\rightarrow -\frac{1}{2}} \bigtriangleup E_1(\phi_K)[s]+ \lim_{s\rightarrow -\frac{1}{2}} \bigtriangleup E_2(\phi_K)[s]  \,\,\, .
\label{zrkc4}
\end{equation}

\section{Zero modes and the heat kernel asymptotic expansion}

\subsection{Conflict at low temperature between zero modes and the heat kernel factorization}

The spectral $K$-heat trace $h_K(\beta)={\rm Tr}_{L^2}\, e^{-\beta K}$ admits an integral kernel representation in terms of the fundamental solution of the $K$-heat equation:
\[
 h_K(\beta)=\int_\Omega \, dx \, K_K(x,x;\beta) \, \, \, \, \,  , \, \, \, \, \, \left(\frac{\partial}{\partial\beta}+K\right)K_K(x,y;\beta)=0 \, \, , \, \, K_K(x,y;0)=\delta(x-y) \, \, .
 \]
The trace, of course, is in the functional sense, $\Omega$ is a finite (but very large) normalization interval on the real line, and $\beta$ is a fictitious inverse temperature. From Sturm-Liouville theory, the eigen-function expansion of the heat kernel
\begin{equation}
K_{K}(x,y;\beta)= f_0^*(y) f_0(x)+\sum_{n=1}^b f_n^*(y)f_n(x) e^{-\beta \omega_n^2}+ \int \! dk \, f_k^{*} (y) \, f_k(x) \, e^{-\beta \omega^2(k)}
\label{integralkernel}
\end{equation}
is founded in solid grounds. Writing (\ref{integralkernel}) we had in mind the general structure of the spectrum of the Schr$\ddot{\rm o}$dinger operator $K$ governing the kink fluctuations: one zero mode $f_0(x)$, a discrete set of bound state eigenfunctions $f_n(x)$, and the continuous spectrum eigenfunctions $f_k(x)$ characterized by a \lq\lq wave number" $k\in\mathbb{R}$.

In the low temperature $\beta\rightarrow +\infty$ regime, only the contribution of the eigenfunction in the algebraic kernel of $K$ survives, whereas the completeness of the eigenfunction expansion fixes the $K$-heat kernel at high-temperature $\beta\rightarrow 0$:
\begin{equation}
\lim_{\beta\rightarrow +\infty} K_K(x,y;\beta)= f_0^*(y) f_0(x) \hspace{1cm},\hspace{1cm} \lim_{\beta\rightarrow 0} K_K(x,y;\beta)= \delta(x-y) \quad .
\label{asympintkernel}
\end{equation}
The standard GDW strategy (and many others) to find the solution of the parabolic equation
\begin{equation}
\left[ \frac{\partial}{\partial \beta}-\frac{\partial^2}{\partial
x^2}+v^2 + V(x) \right] K_{K}(x,y;\beta)=0 \label{heateq}
\end{equation}
with a $\delta$-source on the diagonal of ${\mathbb R}^2$ at infinite temperature $K_{K}(x,y;0)=\delta(x-y)$
is to start from the well known  $K_0$-heat kernel
\[
K_{K_0}(x,y;\beta)=\frac{1}{\sqrt{4\pi \beta\, }} \,
e^{-\beta v^2}\, e^{- \frac{(x-y)^2}{4 \, \beta\,}} \quad , \quad f_k^0(x)=\frac{1}{\sqrt{2 \pi}} e^{ikx} \quad , \quad \omega_0^2(k)=k^2+v^2\quad , \quad k\in{\mathbb R}  \,\,\, ,
\]
and search for a $K$-heat kernel in the factorized form:
\begin{equation}
K_{K}(x,y;\beta)=K_{K_0}(x,y;\beta) \, A(x,y;\beta) \quad , \quad A(x,y;0)=1 \quad .
\label{factorization0}
\end{equation}
The GDW heat kernel expansion
\begin{equation}
K_K(x,y;\beta)=\frac{1}{\sqrt{4\pi\beta}} e^{-\beta v^2}\, e^{- \frac{(x-y)^2}{4 \, \beta\,}} \sum_{n=0}^\infty \, a_n(x,y)\beta^n \label{GDW}
\end{equation}
trades the PDE (\ref{heateq}) by  a recurrence relation between the Seeley densities $a_n(x,y)$. Setting $a_0(x,y)=1$,
one can prove, see e.g. \cite{Roe1988}, that the GDW heat-kernel expansion (\ref{GDW}) is an asymptotic solution
to the PDE equation (\ref{heateq}), i.e., denoting the partial sums in the form $K^{N_t}_K(x,x;\beta)=K_{K_0}(x,x;\beta)\sum_{n=0}^{N_t}\, a_n(x,x)\beta^n$, we have:
\begin{equation}
\left[\frac{\partial}{\partial \beta}-\frac{\partial^2}{\partial
x^2}+v^2 + V(x) \right] K_{K}^{N_t}(x,x;\beta)=\frac{1}{\sqrt{4\pi\beta}}R^{N_t}(x,x;\beta)\beta^{N_t}\, \, \quad , \, \, \quad K_{K}^{N_t}(x,x;0)=\lq\lq \delta(0)" \label{asGDW} \quad ,
\end{equation}
where the remainder $R^{N_t}(x,x;\beta)$ is a $\mathcal{C}^\infty$ function of $x$.

Because
\begin{equation}
\lim_{\beta\rightarrow +\infty} K_{K_0}(x,y;\beta)= 0 \hspace{1cm},\hspace{1cm} \lim_{\beta\rightarrow 0} K_{K_0}(x,y;\beta)= \delta(x-y)  \,\,\, ,
\label{asympintkernel0}
\end{equation}
it is clear that although the asymptotic of $K_K(x,y;\beta)$ read from (\ref{factorization0}) and (\ref{asympintkernel0}) fits with (\ref{asympintkernel}) at high-temperature, it fails at low temperature
due to the zero mode. This discrepancy forces us to limit the integration range in the Mellin transform to a finite one if we use the standard factorization. In other fields, e.g. heat kernel proofs of the index theorem, computation of anomalies in QFT, this is not a difficulty because these effects are related to residua at the poles of the spectral zeta function and the entire parts can be neglected. However, here we are in trouble. The main reason is that there is a value of $\beta$ between low and high temperatures where the heat trace starts to depart from the exact one due to the zero mode. This causes many computational inaccuracies. A considerable amount of ingenuity is needed to obtain good estimations of the one-loop kink mass shift from the kink zeta function via the Mellin transform of the kink heat kernel expansion.

As a loophole, we propose the improved factorization
\begin{equation}
K_{K}(x,y;\beta)=K_{K_0}(x,y;\beta) \, C(x,y;\beta) +g(\beta) e^{-\frac{(x-y)^2}{4\beta}} f_0^*(y) f_0(x)\, \, \, ,
\label{factorization}
\end{equation}
matching the asymptotic behavior of the $K$-heat equation kernel at both low and high temperature, (\ref{asympintkernel}) and (\ref{asympintkernel0}), provided that the $\beta\to \infty$ and $\beta\to 0$ limit of $C(x,y;\beta$ and $g(\beta)$ will be such that:
\begin{equation}
\lim_{\beta \rightarrow 0}C(x,y;0)=1 \hspace{1cm} , \hspace{1cm} \lim_{\beta \rightarrow \infty} g(\beta)=1 \, \, , \, \, \lim_{\beta \rightarrow 0}g(\beta)=0  \,\,\, . \label{heatic2}
\end{equation}

Having modified the standard factorization to adapt the formalism to the existence of zero modes, we follow the GDW route. Plugging the ansatz (\ref{factorization}) into the $K$-heat equation (\ref{heateq}) the following \lq\lq transfer" equation for $C(x,y;\beta)$ arises:
\begin{eqnarray}
&& \left( \frac{\partial}{\partial \beta}+\frac{x-y}{\beta}
\frac{\partial}{\partial x}-\frac{\partial^2}{\partial
x^2}+V(x) \right) C(x,y;\beta) +  \nonumber\\ && + \sqrt{4\pi \beta} \, e^{\beta v^2}\, f_0^*(y) \left[ \frac{dg(\beta)}{d\beta} f_0(x)+ \frac{g(\beta)}{2\beta} f_0(x) +\frac{g(\beta)}{\beta} (x-y) \frac{df_0(x)}{dx} \right] =0 \label{heateq2} \quad .
\end{eqnarray}

\subsection{The modified asymptotic expansion of the kink heat function}
This PDE equation is traditionally solved by means of a power series expansion:
\begin{equation}
C(x,y;\beta) = \sum_{n=0}^\infty c_n(x,y) \, \beta^n  \quad , \quad c_0(x,y)=1 \quad .
\label{expansion}
\end{equation}
Note that $c_0(x,y)=1$ is obligatory to comply with the infinite temperature value of $C(x,y;\beta)$.
Plugging the power expansion in (\ref{heateq2}) this PDE is transformed into the recurrence relations:
\begin{eqnarray}
&&\sum_{n=0}^\infty \left[ (n+1) c_{n+1}(x,y) - \frac{\partial^2 c_n(x,y)}{\partial x^2} +(x-y) \frac{c_{n+1}(x,y)}{\partial x} +V(x) c_n(x,y) \right] \beta^n + \nonumber\\
&&+\sqrt{4\pi \beta} e^{\beta v^2} f_0^*(y) \left[ \frac{dg(\beta)}{d\beta} f_0(x) + \frac{g(\beta)}{2\beta} f_0(x) + (x-y) \frac{g(\beta)}{\beta} \frac{df_0(x)}{dx} \right]=0 \label{relation01} \quad .
\end{eqnarray}
We stress that the modified factorization ansatz leads to new recurrence relations as compared with those encountered by Gilkey and DeWitt: namely, all the terms in the second line of (\ref{relation01}) are new. This structure suggests the optimum choice of $g(\beta)$. Setting
\begin{equation}
\sqrt{\pi\beta} e^{\beta v^2} \frac{dg(\beta)}{d\beta}=v \, \, \, \equiv \, \, \,g(\beta)={\rm Erf}\,(v\sqrt{\beta})  \,\,\, ,
\label{functiong}
\end{equation}
we simplify maximally (\ref{relation01}) in the sense that derivatives of $g(\beta)$ do not enter. Moreover, the error function is selected as the best option to comply simultaneously with (\ref{functiong}) and the asymptotic conditions (\ref{heatic2}).

From the asymptotic expansion of the error function ${\rm Erf}\, z = \frac{2}{\sqrt{\pi}} e^{-z^2} \sum_{n=0}^\infty \frac{2^n}{(2n+1)!!}z^{2n+1}$, the expression (\ref{relation01}) leads to the recurrence relations:
\begin{eqnarray}
&& (n+1) \, c_{n+1}(x,y)-\frac{\partial^2 c_n(x,y)}{\partial
x^2} +(x-y) \frac{\partial c_{n+1}(x,y)}{\partial
x}+V(x) c_n(x,y) + \nonumber \\ && + 2vf_0^*(y)f(x) \delta_{0n} + f_0^*(y)f(x) \frac{2^{n+1}v^{2n+1}}{(2n+1)!!} + (x-y)f_0^*(y) \frac{df_0(x)}{dx} \frac{2^{n+2}v^{2n+1}}{(2n+1)!!}=0 \label{recursive1}\quad .
\end{eqnarray}
The modified heat kernel expansion takes the form:
\begin{equation}
K_{K}(x,y;\beta)=K_{K_0}(x,y;\beta) \,\sum_{n=0}^\infty c_n(x,y) \, \beta^n +{\rm Erf}(\beta) e^{-\frac{(x-y)^2}{4\beta}} f_0^*(y) f_0(x)\quad ,
\label{facto2}
\end{equation}
where the $c_n(x,y)$ densities satisfy the relations (\ref{recursive1}), whereas the diagonal kernel reads:
\[
K_K(x,x,\beta)=\lim_{y\rightarrow x} K_K(x,y,\beta)= \frac{e^{-\beta v^2
}}{\sqrt{4 \pi \beta\,}} \sum_{n=0}^\infty c_n(x,x) \,
\beta^n + {\rm Erf}(\beta) f_0^*(x) f_0(x) \, \, \, , \, \, \, c_n(x,x)=\lim_{y\rightarrow x} c_n(x,y)
\]
and the heat trace is
\[
h_{K}(\beta) =\int_\Omega dx K_K(x,x,\beta) = \frac{e^{-\beta v^2
}}{\sqrt{4 \pi \beta\,}} \sum_{n=0}^\infty c_n(K) \,
\beta^n + {\rm Erf}\,(v\sqrt{\beta}) \quad , \quad c_n(K)=\int_\Omega  dx \, c_n(x,x) \quad .
\]
Thus, to determine the $K$-heat function $h_K(\beta)$ is tantamount to identifying the Seeley coefficients $c_n(K)$. This task is accomplished by solving recursively (\ref{recursive1}) but to do this it is necessary to deal with the following subtlety: the operations of taking the $y\rightarrow x$ limit and the derivatives with respect to $x$ in the formula (\ref{recursive1}) do not commute.
To cope with this problem we introduce the following notation:
\begin{equation}
{^{(k)}C}_n(x)=\lim_{y \rightarrow x} \frac{\partial^k
c_n(x,y)}{\partial x^k} \quad ,
\label{newcoef}
\end{equation}
the Seeley densities, for instance, being: $c_n(x,x)={^{(0)}C}_n(x) $. Moreover, the \lq\lq initial" conditions fix the first coefficient ${^{(k)}C}_0(x)\, , \, \forall k$ to be:
\begin{equation}
{^{(k)} C}_0(x)=\lim_{y\rightarrow x} \frac{\partial^k
c_0}{\partial x^k}= \delta^{k0} \quad .
\label{ini0}
\end{equation}
Taking the $k$-th derivative of the recurrence relation (\ref{recursive1}) with respect to $x$ and later passing to the $y\to x$ limit in the resulting recurrence relations we obtain:
\begin{eqnarray}
{^{(k)} C}_n(x)& =&\frac{1}{n+k} \left[ \rule{0cm}{0.6cm} \right.
{^{(k+2)} C}_{n-1}(x) - \sum_{j=0}^k {k \choose j}
\frac{\partial^j V}{\partial x^j}\, \, {^{(k-j)}
C}_{n-1}(x) - \nonumber \\ &&  -2v f_0(x) \frac{df_0^k(x)}{dx^k} \delta_{0,n-1} - f_0(x) \frac{df_0^k(x)}{dx^k} \frac{2^n v^{2n-1}}{(2n-1)!!} (1+2k) \left. \rule{0cm}{0.6cm} \right] \quad ,
\label{capitalAcoefficients}
\end{eqnarray}
which affords us the identification of ${^{(k)} C}_n(x)$ in a recursive way from the initial ones (\ref{ini0}). Note that the computation of the density $c_n(x,x)={^{(0)} C}_n(x)$ requires the coefficients ${^{(0)} C}_{n-1}(x)$, ${^{(1)} C}_{n-1}(x)$, $\dots$, ${^{(k+2)} C}_{n-1}(x)$, which are in turn determined from the densities ${^{(0)} C}_{n-2}(x)$, ${^{(1)} C}_{n-2}(x)$, $\dots$ ${^{(k+4)} C}_{n-2}(x)$, etcetera, until we reach the densities ${^{(k)} C}_0(x)= \delta^{k0}$.

We list the first three densities $c_n(x,x)$ derived from (\ref{capitalAcoefficients}) for the kink fluctuation operator $K$
\begin{eqnarray}
c_0(x,x)={^{(0)} C}_0(x)&=&1\,\,, \nonumber \\
c_1(x,x)={^{(0)} C}_1(x)&=&-V(x)-4v f_0^2(x)\,\,, \label{Seley1} \\
c_2(x,x)={^{(0)} C}_2(x)&=&-\frac{1}{6} \, \frac{\partial^2 {V}}{\partial
x^2}+\frac{1}{2} \, (V(x))^2 +\frac{4}{3}v^3 f_0^2(x)+4v f_0^2(x) V(x)\,\,, \nonumber
\end{eqnarray}
as well as the first three Seeley coefficients
\begin{eqnarray}
c_0(K)&=&l\,\,, \nonumber\\
c_1(K)&=&- \left< V(x) \right> -4 v\,\,, \label{Seley2} \\
c_2(K)&=&- \frac{1}{6} \left< V''(x) \right> + \frac{1}{2} \left< (V(x))^2 \right> +\frac{4}{3} v^3 + 4 v \left< V(x) f_0^2(x) \right>  \,\, ,\nonumber
\end{eqnarray}
where $l$ is the length of the interval $\Omega$.

Subtraction of the $K_0$-heat trace from the $K$-heat trace expansion amounts to dropping the $c_0(K)$ coefficient, and we find:
\begin{equation}
\overline{h}_K(\beta)=h_{K}(\beta)-h_{K_0}(\beta) = \frac{e^{-\beta v^2
}}{\sqrt{4 \pi}} \sum_{n=1}^\infty c_n(K) \,
\beta^{n-\frac{1}{2}} + {\rm Erf}\,(v\sqrt{\beta})  \,\,\, .
\label{heatfunction3}
\end{equation}
In practical calculations we shall truncate the $\overline{h}_K(\beta)$ series to a finite number of terms{\footnote{
After all, it is an asymptotic series  such that there is an optimum truncation order approximation to the exact value.}}, which defines the function
\begin{equation}
\overline{h}_{K}(\beta;N_t) = \frac{e^{-\beta v^2
}}{\sqrt{4 \pi}} \sum_{n=1}^{N_t} c_n(K) \,
\beta^{n-\frac{1}{2}} + {\rm Erf}\,(v\sqrt{\beta})  \,\,\, .
\label{trunca}
\end{equation}

\subsubsection{An example: the asymptotics of the $\lambda\phi^4$ kink}

To illustrate the power of this modified procedure in the calculation of one-loop kink masses, let us focus on the $\lambda(\phi)_{1+1}^4$ model. $K$ is the Schr$\ddot{\rm o}$dinger operator for the second member of the hierarchy of \lq\lq transparent" P$\ddot{\rm o}$schl-Teller potentials, see \cite{Alonso2002,Alonso2011}. The spectral problem is completely solvable and the $K$-heat trace, a function of $\beta$, is known exactly :
\begin{equation}
\overline{h}_K(\beta)=h_{K}(\beta)-h_{K_0}(\beta) =e^{-3\beta}\, {\rm Erf}(\sqrt{\beta}) + {\rm Erf}(2\sqrt{\beta})
\label{kinkheatfunction} \quad ,
\end{equation}
see \cite{Alonso2002,Alonso2011}. In Figure 1a, the exact $\overline{h}_K(\beta)$ \lq\lq renormalized" heat function has been plotted. A comparison with the approximated heat function given by the truncated asymptotic series $\overline{h}_K(\beta,N_t)$ is shown in Figure 1b for several values of $N_t$. We emphasize that we used the modified GDW heat kernel expansion. The Seeley coefficients needed to compute $\overline{h}_K(\beta,N_t)$ are listed in Table 2 of Section \S.4. The agreement with the exact result is remarkable even for low orders of $N_t$.
\begin{figure}[ht]
\centerline{\includegraphics[height=3cm]{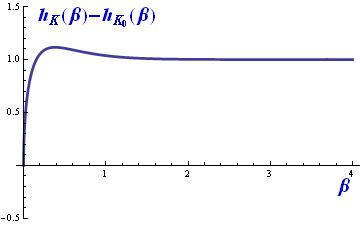}\hspace{1cm}
\includegraphics[height=3cm]{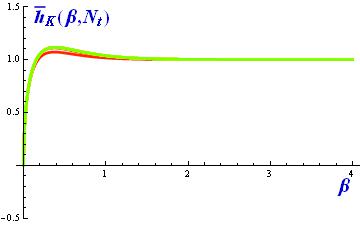} \hspace{1cm}
\includegraphics[height=3cm]{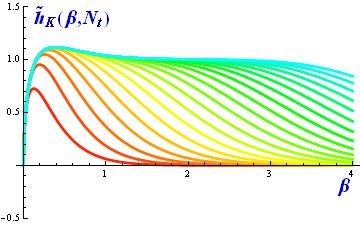}}
\caption{\textit{The exact function $\overline{h}_{K}(\beta)$ (left), partial sums $\overline{h}_{K}(\beta,N_t)$ from the modified factorization method (middle) and partial sums $\widetilde{h}_{K}(\beta,N_t)$ from the standard GDW factorization method (right) for the $\lambda(\phi)_{1+1}^4$ kink.}}
\end{figure}
In Figure 1c different plots of $\widetilde{h}_{K}(\beta;N_t)$, the truncated heat functions derived from the GDW  factorization (\ref{factorization0}), are shown as colored lines for increasing values of $N_t$ from the bottom (red) to the top (violet):
\begin{equation}
\widetilde{h}_{K}(\beta;N_t) = \frac{e^{-\beta v^2
}}{\sqrt{4 \pi}} \sum_{n=1}^{N_t} {a}_n(K) \,
\beta^{n-\frac{1}{2}} \qquad .
\label{trunca2}
\end{equation}
The Seeley coefficients $a_n(K)$ for the $\lambda\phi^4$ kink are taken from the References \cite{Alonso2002,Alonso2011}. The main point is perfectly clear here:
\[
\lim_{\beta \rightarrow \infty} \overline{h}_{K}(\beta)=1 = \lim_{\beta \rightarrow \infty} \overline{h}_{K}(\beta,N_t) \, \, \, , \, \, \, \forall N_t \qquad \quad , \qquad \quad \lim_{\beta \rightarrow \infty} \widetilde{h}_{K}(\beta,N_t)=0 \qquad .
\]
The exact heat trace and the asymptotic approximation through the modified GDW factorization for the $\lambda\phi^4$
kink tend to unity when the temperature decreases to zero: $\beta=\infty$.  The asymptotic approximation from the standard GDW factorization, however, tends to zero at low temperature. It is clear that the zero mode is not captured in this latter case and one must choose an optimum value of $\beta$ as the upper limit of the integral in the Mellin transform. There is a second point to be emphasized: $\overline{h}_{K}(\beta,N_t)$ approaches $\overline{h}_{K}(\beta)$
much faster than $\widetilde{h}_{K}(\beta,N_t)$. Higher $N_t$ is necessary to fit in with the exact function in this latter case. Because $\zeta_K(s)-\zeta_{K_0}(s)=\frac{1}{\Gamma(s)}\int_0^\infty \, d\beta \, \beta^{s-1} \overline{h}_K(\beta)$, it is evident that a better approximation to the \lq\lq renormalized" kink zeta function is $\zeta_K(s)-\zeta_{K_0}(s)=\frac{1}{\Gamma(s)}\int_0^\infty \, d\beta \, \beta^{s-1} \overline{h}_K(\beta;N_t)$ rather than $\zeta_K(s)-\zeta_{K_0}(s)=\frac{1}{\Gamma(s)}\int_0^\infty \, d\beta \, \beta^{s-1} \widetilde{h}_K(\beta,N_t)$.

\subsection{Kink mass quantum corrections in one-component scalar field theory}

The Mellin transform of the renormalized $K$-heat trace expansion $\overline{h}_K(\beta)$ (\ref{heatfunction3})
gives us the renormalized $K$-zeta function as the series (involving Euler Gamma functions)
\[
\zeta_K(s)-\zeta_{K_0}(s)=\frac{1}{\sqrt{4\pi}} \sum_{n=1}^\infty c_n(K) \frac{\Gamma[s+n-\frac{1}{2}]}{\Gamma[s]} -\frac{1}{\sqrt{\pi}} v^{-2s} \frac{\Gamma[s+\frac{1}{2}]}{s\Gamma[s]} \quad ,
\]
which in turn, recall (\ref{zrkc}), prompts the following expression for the regularized kink Casimir energy $\bigtriangleup E_1(\phi_K)[s]$:
\begin{eqnarray*}
\bigtriangleup E_1(\phi_K)[s]&=& \frac{\hbar \gamma_d^2}{2} \left( \frac{\mu^2}{m_d^2} \right)^{s+\frac{1}{2}} \Big[ -\frac{1}{\sqrt{4\pi}} \frac{\left<V(x)\right>}{v^{1+2s}} \frac{\Gamma[s+\frac{1}{2}]}{\Gamma[s]} - \frac{2}{\sqrt{\pi}\,v^{2s} } \frac{\Gamma[s+\frac{1}{2}]}{\Gamma[s]} + \\ &+& \frac{1}{\sqrt{4\pi}} \sum_{n=2}^\infty \frac{c_n(K)}{v^{2n+2s-1}} \frac{\Gamma[s+n-\frac{1}{2}]}{\Gamma[s]} - \frac{1}{\sqrt{\pi}v^{2s}} \frac{\Gamma[s+\frac{1}{2}]}{s\Gamma[s]} \, \Big] \quad .
\end{eqnarray*}
Here we have used the explicit expression of $c_1(K)$ given in (\ref{Seley2}) to write the contribution of the first term of the series. Because the contribution of the regularized one-loop mass counter-term $\bigtriangleup E_2(\phi_K)[s]$, see (\ref{zrkc3}), is
\[
\bigtriangleup E_2(\phi_K)[s]=\frac{\hbar\gamma_d^2}{2}\left(\frac{\mu^2}{m_d^2}\right)^{s+\frac{1}{2}}
\frac{\langle V(x) \rangle}{\sqrt{4\pi}}\frac{\Gamma[s+\frac{1}{2}]}{\Gamma[s]}\frac{1}{v^{2s+1}}  \,\,\, ,
\]
a crucial cancelation, obeying the heat kernel renormalization criterion, see e.g. \cite{Bordag2002}, occurs after the addition of these two regularized quantities and we obtain the following series for the regularized one-loop kink mass shift $\bigtriangleup E(\phi_K)[s]=\bigtriangleup E_1(\phi_K)[s]+\bigtriangleup E_2(\phi_K)[s]$:
\[
\frac{\bigtriangleup E(\phi_K)[s]}{\hbar \gamma_d^2}=\frac{1}{2} \left( \frac{\mu^2}{m_d^2} \right)^{s+\frac{1}{2}} \left[  \frac{1}{\sqrt{4\pi}} \sum_{n=2}^\infty \frac{c_n(K)}{v^{2n+2s-1}} \frac{\Gamma[s+n-\frac{1}{2}]}{\Gamma[s]} - \frac{2}{\sqrt{\pi}\,v^{2s} } \frac{\Gamma[s+\frac{1}{2}]}{\Gamma[s]} - \frac{1}{\sqrt{\pi}v^{2s}} \frac{\Gamma[s+\frac{1}{2}]}{s\Gamma[s]} \, \right]  \,\,\, .
\]
Still, the last two terms are divergent at the physical point $s=-\frac{1}{2}$, which is a pole of $\Gamma(s+\frac{1}{2})$. There is, however, a miraculous cancelation that is observed by taking the dangerous
limit carefully:
\begin{eqnarray*}
&& \lim_{s\rightarrow -\frac{1}{2}} \left[ - \frac{2}{\sqrt{\pi}\,v^{2s} } \frac{\Gamma[s+\frac{1}{2}]}{\Gamma[s]} - \frac{1}{\sqrt{\pi}v^{2s}} \frac{\Gamma[s+\frac{1}{2}]}{s\Gamma[s]} \right] = \lim_{\varepsilon\rightarrow 0} \Big[ \frac{v}{\pi} \left(\frac{1}{\varepsilon} -\,[\gamma +2 \log v+ \psi(-{\textstyle\frac{1}{2}})] \right)+ {\cal O}_1(\varepsilon)- \\
&& -\frac{v}{\pi} \left(\frac{1}{\varepsilon} -[\gamma +2 \log v+ \psi(-{\textstyle\frac{1}{2}})]+2 \right)+ {\cal O}_2(\varepsilon) \Big] = -\frac{2v}{\pi} \quad.
\end{eqnarray*}
Thus, zero point and mass renormalization get rid of all the divergences
and we end with the renormalized one-loop mass shift formula derived from the modified asymptotic series expansion of the $K$-heat function:
\begin{equation}
\frac{\bigtriangleup E(\phi_K)}{\hbar \gamma_d^2}=-\frac{v}{\pi}  -\frac{1}{8\pi} \sum_{n=2}^\infty c_n(K)(v^2)^{1-n}\Gamma[n-1]  \,\,\, .
\label{computation1}
\end{equation}
We have elaborated a Mathematica code, shown in the Appendix of this paper, which automatizes the computation of the quantum correction to the kink mass in one-component scalar field theory models. The algorithm is based on the evaluation of the $N_t$-th order partial sum of the series (\ref{computation1}),
\begin{equation}
\frac{\bigtriangleup E(\phi_K;N_t)}{\hbar \gamma_d^2}=-\frac{v}{\pi}  -\frac{1}{8\pi} \sum_{n=2}^{N_t} c_n(K)(v^2)^{1-n}\Gamma[n-1]
\label{numericalcorrection}
\end{equation}
as a good estimation of the quantum correction. This Mathematica program uses the recurrence relations (\ref{capitalAcoefficients}) and the definition of the Seeley coefficients to figure out the coefficients $c_n(K)$ in (\ref{numericalcorrection}). The inputs needed in this program only include the potential term $U(\phi)$, which characterizes in action (\ref{action}) the specific model chosen, the two vacua which are connected by the kink solution whose mass quantum correction we are interested in, and the order of truncation $N_t$ to be considered in our estimations (\ref{numericalcorrection}).

\section{Specific $(1+1)$-D one-component scalar field theory models}

In this Section we shall apply formula (\ref{numericalcorrection}), in the environment of the Mathematica code, to compute the one-loop kink mass shift in several well selected (1+1)-dimensional scalar field theory models. We start with the very well known sine-Gordon and $\phi^4$ kinks as a good test of the new procedure. We shall see that in these paradigmatic cases the new method works extremely well whereas the analysis helps in the conceptual understanding of the modified heat kernel expansion. We shall then move on to explore more exotic avenues: First, we shall apply the procedure to compute the kink mass quantum correction to the $\sigma=N$ \lq\lq parent" models, see \cite{Alonso2012}, a very favorable arena because the second-order fluctuation operator $K=N^2-\frac{N(N+1)}{{\rm cosh}^2x}, \, N\in{\mathbb N}$  belongs to the transparent P$\ddot{\rm o}$sch-Teller hierarchy. Second, a one-parametric family of generalized $\phi^6$ models having kinks and \lq\lq double" kinks, see \cite{Alonso2011},  will be also addressed and the one-loop kink mass shifts estimated. Here, to calibrate the quality of the results only analyze of self-consistency are available. The $K$ operator does not belong to any class of Sch$\ddot{\rm o}$dinger operators with known spectra. Therefore, no other procedures of computation such as the DHN or the Comtet-Cahill-Glauber formula are applicable. Third, the kinks in the double sine-Gordon model \cite{Mussardo} will be also considered. Again, we shall obtain estimations of the kink mass quantum shifts that are not accessible with other procedures.

\subsection{The sine-Gordon model}

The part of the energy density potential independent of the field spatial derivatives is in the sine-Gordon model:
\[
U(\phi)=1-\cos \phi \quad .
\]
The sine-Gordon kink $\phi_K(x)=4\arctan e^x + 2\pi n$, $n\in \mathbb{Z}$, is a soliton solution that connects the vacua $2\pi n$ and $2\pi (n+1)$. The second-order small vacuum and kink fluctuation operators are respectively:
\[
K_0=-\frac{d^2}{dx^2}+1 \hspace{1cm},\hspace{1cm} K=-\frac{d^2}{dx^2}+1-2\,{\rm sech}^2 x \quad .
\]
Thus, $v=1$ and $V(x)=-2\,{\rm sech}^2 x$, in agreement with the notation used in the previous Sections. The spatial derivative of the kink solution is the (normalized) zero mode of  fluctuation: $f_0(x)=\frac{1}{\sqrt{2}} \,{\rm sech}\,x$. From (\ref{capitalAcoefficients}) and spatial integration of the Seeley densities we find that all the Seeley coefficients for the sine-Gordon kink vanish. This reveals the hidden nature of the modified heat kernel expansion: the choice $g(\beta)={\rm Erf}\sqrt{\beta}$ means that the new expansion is with respect to the heat
kernel of the sine-Gordon kink, not of a constant background!!. Therefore, $\overline{h}_K(\beta)={\rm Erf}(\sqrt{\beta})$ is the exact sine-Gordon kink heat trace. The numerical estimation of the sine-Gordon kink mass quantum correction through (\ref{numericalcorrection}) thus provides the exact result: $\frac{\Delta E(\phi_K)}{\hbar \gamma_d^2}=-\frac{1}{\pi}=-0.31831$. The optimum number of terms in the asymptotic formula is $N_t=0$ !! in the always surprising sine-Gordon model.

\subsection{The $\lambda(\phi)_{1+1}^4$ model}

The potential term in this model is:
\[
U(\phi)=\frac{1}{2}(\phi^2-1)^2 \quad .
\]
The kink solitary wave $\phi(x)=\tanh x$ connects the two vacua $\phi^{(1)}=-1$ and $\phi^{(2)}=1$. The vacuum and kink Hessian operators are respectively:
\[
K_0=-\frac{d^2}{dx^2}+4 \hspace{1cm},\hspace{1cm} K=-\frac{d^2}{dx^2}+4-6\,{\rm sech}^2 x \quad ,
\]
such that $v=2$ and $V(x)=-6\,{\rm sech}^2 x$. The zero mode is the spatial derivative of the kink, i.e.,
properly normalized, $f_0(x)=\frac{\sqrt{3}}{2} \,{\rm sech}^2\,x$. The symbolic resolution of the recurrence relations (\ref{capitalAcoefficients}) for the analytic expressions of $V(x)$ and $f_0(x)$ in  the Mathematica code provides  Seeley densities that, integrated over the whole real line, become the Seeley coefficients entered into the formula (\ref{numericalcorrection}). These coefficients are listed in the Table 1. In this table we have also specified the values of the one-loop mass shift obtained from the modified asymptotic series (\ref{numericalcorrection}) for different values of the truncation order, $N_t$, graphically represented by the blue line of the attached figure. For the sake of the comparison we also collect from \cite{Alonso2002,Alonso2011} the same data obtained from the standard asymptotic series, represented by the red line.

\begin{table}[ht]
\centerline{\begin{tabular}{|c||c|c|}
\hline
\multicolumn{2}{|c|}{{\small \it Kink Seeley Coefficients}} \\ \hline
$n$ & $c_n(K)$  \\ \hline
$1$  & $4.00000$ \\
$2$  & $2.66667$ \\
$3$  & $1.06667$ \\
$4$  & $0.304762$ \\
$5$  & $0.0677249$ \\
$6$  & $0.0123136$ \\
$7$  & $0.0018944$ \\
$8$  & $0.000252587$ \\
$9$  & $0.0000297161$ \\
$10$  & $3.12801 \cdot 10^{-6}$ \\ \hline
\end{tabular}\hspace{0.4cm} \begin{tabular}{|c||c|c|}
\hline
\multicolumn{2}{|c|}{{\small \it Kink Mass Shift Estimation}} \\ \hline
$N_t$ & ${\bigtriangleup {E}(\phi_K;N_t)}/{\hbar \gamma_d^2}$  \\ \hline
 - & - \\
$2$ &  $-0.663146$ \\
$3$ &  $-0.665798$ \\
$4$ &  $-0.666177$ \\
$5$ &  $-0.666240$ \\
$6$ &  $-0.666252$ \\
$7$ &  $-0.666254$ \\
$8$ &  $-0.666254$ \\
$9$ &  $-0.666255$ \\
$10$ &  $-0.666255$ \\ \hline
\end{tabular}\hspace{0.4cm} \begin{tabular}{c}
\includegraphics[height=4.5cm]{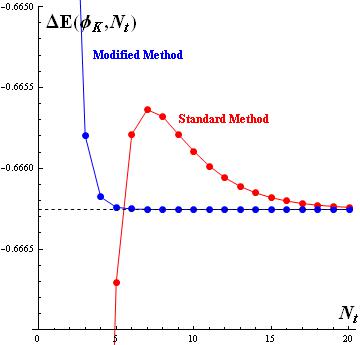}
\end{tabular}}
\caption{\textit{Seeley coefficients (left), kink mass shift estimation (middle) and comparison between the shifts  obtained from the modified (blue) and from the standard (red) asymptotic series (right) for several truncation orders
in the $\phi^4$ model.}}
\end{table}

In sum, the one-loop mass shift obtained from the modified heat trace expansion truncated at the order $N_t=10$
is:
\[
\frac{\Delta E(\phi_K,N_t=10)}{\hbar \gamma_d^2}=-0.666255 \quad .
\]
This estimation very precisely reproduces the exact result $\frac{\Delta E(\phi_K)}{\hbar \gamma_d^2}=-\frac{1}{\pi}(3-\sqrt{3}{\rm arccos}\frac{\sqrt{3}}{2})=\frac{1}{2\sqrt{3}}-\frac{3}{\pi}=-0.666255$ found in this model, see \cite{Alonso2002,Alonso2011}.

Three points are worthy of mention:
\begin{enumerate}

\item The Seeley coefficients for the $\phi^4$ kink obtained through the modified heat trace expansion are exactly the same as the Seeley coefficients for the sine-Gordon kink obtained from the standard heat trace asymptotics.

 \item  It is clear from the above Figure that the convergence obtained upon using the modified approach is much better than in the standard procedure. The optimum asymptotic value is reached at a truncation order of three
     or four in the new method, whereas in the traditional approach more than ten terms were necessary.

  \item  All this stresses how a better approximation to the $K$-heat function is obtained in the modified approach. The distortions introduced by the zero mode are almost suppressed when the whole effect of the
      kink fluctuations is calibrated with respect to the simplest kink with a zero mode !!
\end{enumerate}

\subsection{The $\sigma=3$ and $\sigma=4$ parent potential models}

In reference \cite{Alonso2012}, a family of (1+1)-dimensional scalar field models governed by the action (\ref{action}) with the following \lq\lq parent" potentials{\footnote{In \cite{Alonso2012} it is explained that these models are well defined only for $|\phi|\leq 1$.}}
\begin{equation}
U^{(\sigma)}(\phi)=\frac{2 \Gamma[\frac{1}{2}+\frac{\sigma}{2}]^2}{\pi \Gamma[\frac{\sigma}{2}]^2} \left(1- I^{-1}\textstyle \left[|\phi|;\frac{1}{2},\frac{\sigma}{2} \right]\right)^\sigma
\label{potentialinPhi}
\end{equation}
is fully analyzed. $I^{-1}(z;a,b)$ is the inverse incomplete regularized Beta function and the reason for the denomination \lq\lq parent" is that the potential is identified by an inverse procedure. It is mandatory that the $K$-operator be
\[
K=-\frac{d^2}{dx^2}+\sigma^2-\frac{\sigma(\sigma +1)}{{\rm cosh}^2x} \quad ,
\]
whereas the kink profile is identified from the zero mode of $K$:
\begin{equation}
\phi_K^{(\sigma)}(\overline{x})= {\rm sign}(\overline{x})\,\, I[\textstyle \tanh^2 \overline{x};\frac{1}{2},\frac{\sigma}{2}]  \,\,\, .
\label{kinkN2}
\end{equation}
It is not completely clear from (\ref{potentialinPhi}) and/or (\ref{kinkN2}), but $\sigma=1$ and $\sigma=2$ respectively correspond to the sine-Gordon and $\lambda(\phi)_{1+1}^4$ models. Here we shall study the next two cases in depth (for $\sigma$ a positive integer): $\sigma=3$ and $\sigma=4$.

In the first case, $\sigma=3$, the small vacuum and kink fluctuation operators are
\[
K_0=-\frac{d^2}{dx^2}+9 \hspace{1cm},\hspace{1cm} K=-\frac{d^2}{dx^2}+9-12\,{\rm sech}^2 x \quad .
\]
Thus, we have $v=3$ and $V(x)=-12\,{\rm sech}^2 x$ and the normalized zero mode is: $f_0(x)=\frac{\sqrt{15}}{4} \,{\rm sech}^3\,x$. The Seeley coefficients listed in the Table 2 are obtained from (\ref{capitalAcoefficients}). We also specify the partial sums $\overline{h}_K(\beta,N_t)$ (\ref{numericalcorrection}) for distinct values of the truncation order $N_t$. The values of the one-loop mass shifts obtained from these partial sums are depicted in the attached figure by the blue line and compared with the data obtained in \cite{Alonso2012}) using the standard procedure, which are represented by the red line.

\begin{table}[ht]
\centerline{\begin{tabular}{|c||c|}
\hline
\multicolumn{2}{|c|}{{\small \it Kink Seeley Coefficients}} \\ \hline
$n$  & $c_n(K)$  \\ \hline
$1$  &  $12.0000$  \\
$2$  & $24.0000$  \\
$3$  & $35.2000$  \\
$4$  & $39.3143$  \\
$5$  & $34.7429$  \\
$6$  & $25.2306$  \\
$7$  & $16.5208$  \\
$8$  & $8.27770$  \\
$9$  & $3.89498$  \\
$10$ & $1.63998$  \\ \hline
\end{tabular}\hspace{0.4cm} \begin{tabular}{|c||c|}
\hline
\multicolumn{2}{|c|}{{\small \it Kink Mass Shift Estimation}} \\ \hline
$N_t$ & ${\bigtriangleup {E}(\phi_K;N_t)}/{\hbar \gamma_d^2}$  \\ \hline
- & - \\
$2$  & $-1.06103$ \\
$3$  & $-1.07832$ \\
$4$  & $-1.08262$ \\
$5$  & $-1.08388$ \\
$6$  & $-1.08429$ \\
$7$  & $-1.08443$ \\
$8$  & $-1.08448$ \\
$9$  & $-1.08449$ \\
$10$ & $-1.08450$ \\ \hline
\end{tabular}\hspace{0.4cm} \begin{tabular}{c}
\includegraphics[height=4.5cm]{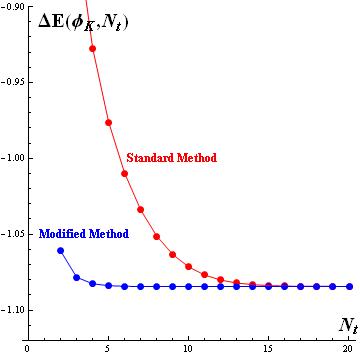}
\end{tabular}}
\caption{\textit{Seeley coefficients (left), kink mass shift estimation (middle) and comparison between the shifts obtained from the modified (blue) and from the standard (red) asymptotic series (right) in the $\sigma=3$ parent potential model.}}
\end{table}

From the data in Table 2, the one-loop kink mass quantum correction is estimated to be
\[
\frac{\Delta E(\phi_K,N_t=10)}{\hbar \gamma_d^2}=-1.08450  \,\,\, .
\]
In \cite{Alonso2012} the exact kink mass quantum correction in this model is written in terms of the bound state energies and the threshold of the continuous spectrum of the $K$ operator: $\frac{\Delta E(\phi_K)}{\hbar \gamma_d^2}=-\frac{1}{\pi}(6-\sqrt{5}{\rm arccos}\frac{\sqrt{5}}{3}-\sqrt{8}{\rm arccos}\frac{\sqrt{8}}{3})=-1.08451$. Therefore, the estimation obtained with the modified method is very precise. In Figure 2a we plot the renormalized kink heat functions $\overline{h}_K(\beta)$ for this model. In Figure 2b we depict the partial sums $\overline{h}_K(\beta,N_t)$ (\ref{trunca}) derived from the modified asymptotic series for several values of $N_t$. Finally, in Figure 2c we represent the partial sums $\widetilde{h}_K(\beta,N_t)$ (\ref{trunca2}), calculated by means of the GDW standard procedure.

\begin{figure}[ht]
\centerline{\includegraphics[height=3cm]{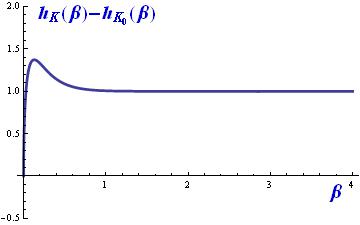}\hspace{1cm}
\includegraphics[height=3cm]{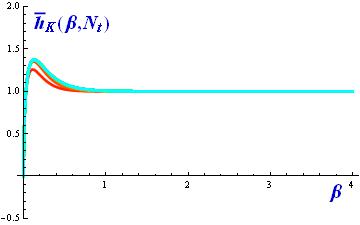} \hspace{1cm}
\includegraphics[height=3cm]{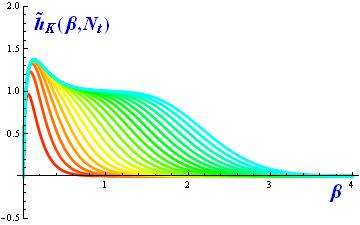}}
\caption{\textit{The exact function $\overline{h}_{K}(\beta)$ (left), partial sums $\overline{h}_{K}(\beta,N_t)$ from the modified factorization (middle) and partial sums $\widetilde{h}_{K}(\beta,N_t)$ from the standard factorization (right) in the $\sigma=3$ parent potential model.}}
\end{figure}

In the second case $\sigma=4$, the small vacuum and kink fluctuation operators are
\[
K_0=-\frac{d^2}{dx^2}+16 \hspace{1cm},\hspace{1cm} K=-\frac{d^2}{dx^2}+16-20\,{\rm sech}^2 x \quad .
\]
Thus, we have $v=4$ and $V(x)=-20\,{\rm sech}^2 x$ and the normalized zero mode is: $f_0(x)=\frac{\sqrt{35}}{4\sqrt{2}} \,{\rm sech}^4\,x$. In Table 3 we summarize the results obtained for this model showing the Seeley coefficients and the quantum mass shift estimations for different values of the truncation order $N_t$.

\begin{table}[ht]
\centerline{\begin{tabular}{|c||c|}
\hline
\multicolumn{2}{|c|}{{\small \it Kink Seeley Coefficients}} \\ \hline
$n$  & $c_n(K)$  \\ \hline
$1$  & $24.0000$  \\
$2$  & $96.0000$  \\
$3$  & $294.200$  \\
$4$  & $705.829$  \\
$5$  & $1367.77$  \\
$6$  & $2206.55$  \\
$7$  & $3035.81$  \\
$8$  & $3632.62$  \\
$9$  & $3841.44$  \\
$10$ & $3637.21$  \\ \hline
\end{tabular}\hspace{0.4cm} \begin{tabular}{|c||c|}
\hline
\multicolumn{2}{|c|}{{\small \it Kink Mass Shift Estimation}} \\ \hline
$N_t$ & ${\bigtriangleup {E}(\phi_K;N_t)}/{\hbar \gamma_d^2}$  \\ \hline
- & - \\
$2$  & $-1.51197$ \\
$3$  & $-1.55773$ \\
$4$  & $-1.57144$ \\
$5$  & $-1.57642$ \\
$6$  & $-1.57843$ \\
$7$  & $-1.57930$ \\
$8$  & $-1.57969$ \\
$9$  & $-1.57986$ \\
$10$ & $-1.57995$ \\ \hline
\end{tabular}\hspace{0.4cm} \begin{tabular}{c}
\includegraphics[height=4.5cm]{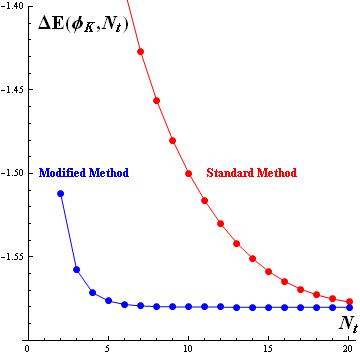}
\end{tabular}}
\caption{\textit{Seeley coefficients (left), kink mass shift estimation (middle) and comparison between the shifts obtained from the modified (blue) and from the standard (red) asymptotic series (right) in the $\sigma=4$ parent potential model.}}
\end{table}

The one-loop kink mass quantum correction is estimated to be
\[
\frac{\Delta E(\phi_K,N_t=14)}{\hbar \gamma_d^2}=-1.58003  \,\,\, .
\]
As in the previous example, we conclude that this result is very precise, recalling that the exact result is: $\frac{\Delta E(\phi_K)}{\hbar \gamma_d^2}=-\frac{1}{\pi}(10-\frac{\pi}{\sqrt{3}}- \sqrt{15}\arcsin\frac{1}{4}-\sqrt{7} \arcsin \frac{3}{4})=-1.58003$, see \cite{Alonso2012}.

Taken together, the results shown in the Tables 2 and 3 and represented in the above Figures confirm the pattern found in the $\lambda(\phi)^4_{1+1}$ model: (1) $\overline{h}_K(\beta,N_t)$ matches the exact $\overline{h}_K(\beta)$ at both small and large $\beta$. $\widetilde{h}_K(\beta,N_t)$, however, only behaves like the exact heat trace at high $\frac{1}{\beta}$ but, even though the approximation is better with increasing $N_t$, there is always a finitely large value of $\beta$, such that $\widetilde{h}_K(\beta,N_t)$ starts to tend to $0$ instead of going to $1$. (2) A much faster convergence to the optimum estimation is achieved using the modified GDW heat kernel factorization. (3) Some hierarchical structure arises: the Seeley coefficients for the kink heat trace modified expansion in the $\sigma$ parent potential model are those found in the $\sigma-1$ model coming from the standard GDW factorization. This third point suggests that one could use any operator
\[
K_N=-\frac{d^2}{dx^2}+N^2-\frac{N(N+1)}{{\rm cosh}^2 x} \quad , \quad N=0,1,2,3, \dots
\]
in the transparent P$\ddot{\rm o}$sch-Teller hierarchy as the starting point of the heat kernel expansion for any differential operator with a zero mode.

\subsection{A family of $(\phi)_{1+1}^6$ models}

We now consider a one-parametric family of $\phi^6$ models. The potential, depending on the coupling constant $a$, is:
\[
U(\phi;a) = \frac{1}{2} (\phi^2 + a^2) (\phi^2-1)^2 \quad ,
\]
see Figure 3a. The vacuum orbit is ${\cal M}=\{-1,1\}$ and the kinks that connect these vacuum points are:
\begin{equation}
\phi_K(x;a)= \frac{a (-1+e^{2\sqrt{1+a^2}\,x})}{\sqrt{4e^{2\sqrt{1+a^2}\,x}+a^2 (1+e^{2\sqrt{1+a^2}\,x})^2}} \quad . \label{solutionphi6}
\end{equation}
In Figure 3b some members of this kink family have been depicted together with their energy densities. In these graphics it is clear that this model presents double kink solutions, a character accentuated when the parameter $a$ approaches zero. Indeed, at the $a\rightarrow 0$ limit the potential reduces to $U(\phi,0)=\frac{1}{2} \phi^2 (\phi^2-1)^2$, i.e., the pure $\phi^6$ Lohe/Khare model \cite{Lohe}; a new vacuum point $\phi=0$ arises, and the vacuum orbit becomes: ${\cal M}=\{-1,0,1\}$. The $\lim_{a\to 0}\phi_K(x;a)$ is a configuration where
the $\phi_K(a)$ kink solution splits into two single kinks, one connecting the vacua $\phi=-1$ and $\phi=0$ and the other connecting $\phi=0$ and $\phi=1$.

\begin{figure}[h]
\centerline{\includegraphics[height=2.5cm]{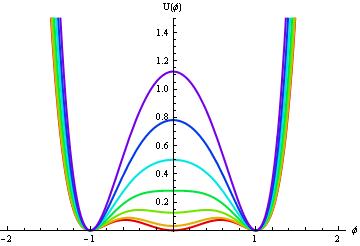}\hspace{0.7cm} \includegraphics[height=2.5cm]{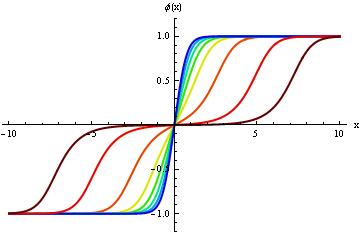} \hspace{0.7cm} \includegraphics[height=2.5cm]{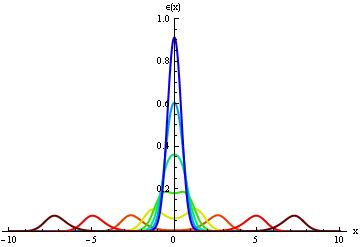}}
\caption{\small \textit{Graphical representation of the potential (left), several kink profiles  (middle) and kink energy densities (right) for $a=0.001,0.01,0.25,0.5,0.75,1,1.25,1.5$.}}
\end{figure}

The particle masses and the kink potential wells in this family of models are
\[
v^2=4(1+a^2) \hspace{0.3cm}, \hspace{0.3cm} V(x;a)=\frac{15 (4 a+1)^2}{\left[2 a \cosh \left(2 \sqrt{a^2+1} x\right)+2 a+1\right]^2}-\frac{6 \left(a^2+3\right) (4 a+1)}{2 a \cosh \left(2 \sqrt{a^2+1} x\right)+2 a+1} \quad .
\]
The Seeley coefficients of the modified asymptotic series evaluated using (\ref{capitalAcoefficients}) have been implemented in the Mathematica code available in this paper for several selected values of $a$. Thus, the one-loop kink mass shifts are computed and are shown in Table 4 with a truncation of $N_t$ terms. All these data are represented graphically in the attached figure in Table 4.

\vspace{0.2cm}

\begin{table}[ht]
\begin{tabular}{|c|c|c|}
\hline
{\small $a$} & {\small ${\bigtriangleup {E}(\phi_K;N_t)}/{\hbar \gamma_d^2}$} & {\small $N_t$} \\ \hline
0.001 & -1.95321 & 9 \\ \hline
0.01 & -1.65859 & 9 \\ \hline
0.05 & -1.44956 & 9 \\ \hline
0.1 & -1.35308 & 9 \\ \hline
0.2 & -1.24166 & 9 \\ \hline
0.3 & -1.17150  & 10\\ \hline
0.4 & -1.11564  & 10\\ \hline
0.5 & -1.07925  & 11 \\ \hline
0.6 & -1.05903  & 11 \\ \hline
\end{tabular} \hspace{0.3cm}
\begin{tabular}{|c|c|c|}
\hline
{\small $a$} & {\small ${\bigtriangleup {E}(\phi_K;N_t)}/{\hbar \gamma_d^2}$} & {\small $N_t$}  \\ \hline
0.7 & -1.05493  & 11 \\ \hline
0.8 & -1.06099  & 11 \\ \hline
0.9 & -1.07728  & 11 \\ \hline
1.0 & -1.10137 & 11  \\ \hline
1.1 & -1.13185  & 11 \\ \hline
1.2 & -1.16744  & 11 \\ \hline
1.3 & -1.20727  & 11 \\ \hline
1.4 & -1.25054  & 10 \\ \hline
1.5 & -1.29667 & 9 \\ \hline
\end{tabular} \hspace{0.3cm}
\begin{tabular}{c}
\includegraphics[height=3.5cm]{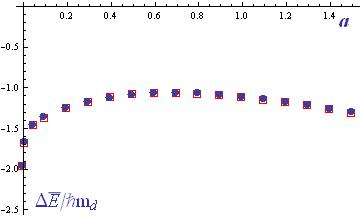}
\end{tabular}
\caption{\textit{Quantum corrections to the kink masses in the $\phi^6$ family for several values of the parameter $a$ (left) and their graphical representation (right). The red squares represent the data obtained in the standard GDW approach, whereas the blue dots are the new data coming from the modified factorization.}}
\end{table}

We observe in Table 4 that the quantum correction tends to infinity as $a$ approaches zero. The reason for this stems from the fact that the vacua that are connected by the kink in the case $a=0$ give rise to different meson masses: $\frac{\partial^2 U}{\partial \phi^2}[1]=4$ and $\frac{\partial^2 U}{\partial \phi^2}[0]=1$, see \cite{Alonso2002} and references therein to see a treatment of this problem.

\subsection{The double sine-Gordon model family}

Finally, we shall apply the procedure to a family of models characterized by the action (\ref{action}) where
the potential is:
\[
U(\phi;a) = 1-(1-a)\cos \phi-a \cos (2\phi) \quad , \quad 0\leq a \leq 1 \, \, , \, \, a\in {\mathbb R} \quad .
\]
These models (here we are considering $a$ as a free parameter) are referred to as the double sine-Gordon model in the literature, see \cite{Mussardo}. When $a$ varies in the $[0,1]$ range, $U(\phi;a)$ interpolates between the sine-Gordon, $a=0$, and the re-scaled sine-Gordon, $a=1$, potentials:
\[
U(\phi;0)= 1-\cos \phi =U_{\rm sG}^{(2\pi)}(\phi) \hspace{1cm},\hspace{1cm} U(\phi;1)= 1-\cos 2\phi =U_{\rm sG}^{(\pi)}(\phi)  \,\,\, .
\]
If $0\leq a < 1$, the vacuum orbit is ${\cal M}=\{2\pi k\}$, $k\in \mathbb{Z}$. When $a$ becomes greater than $0$ but is strictly less than $1$, new relative minima (false vacua) emerge between each consecutive pair of vacua. At $a=1$, the new minima become absolute minima, the vacuum points of the re-scaled sine-Gordon model, see Figure 4 (left), such that the vacuum orbit becomes ${\cal M}=\{\pi k\}$, $k\in \mathbb{Z}$.
The kink solution of each member of the family with $a$ strictly less than $1$ is:
\begin{equation}
\phi_K(x;a)= -2 \arctan \frac{\sqrt{1+3a}}{\sqrt{1-a} \sinh (\sqrt{1+3a}x)} = \pi + 2 \arctan \frac{\sqrt{1-a} \sinh (\sqrt{1+3a}x)}{\sqrt{1+3a}}  \,\,\, .
\label{mixedsoliton}
\end{equation}
In Figure 4 we depict the kink profiles for several values of the coupling constant together with their energy densities.

\begin{figure}[h]
\centerline{\includegraphics[height=2.5cm]{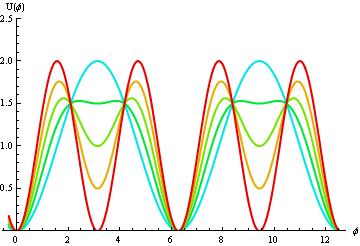}\hspace{0.6cm} \includegraphics[height=2.5cm]{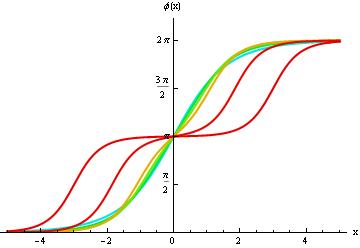}\hspace{0.6cm} \includegraphics[height=2.5cm]{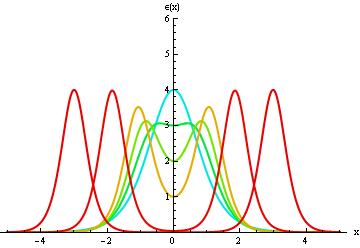}}
\caption{\small \textit{Graphical representation of the family of potentials for $a=0,0.25,0.5,0.75,1$ (left), kink solution profiles (middle), kink energy densities (right) for $a=0,0.25,0.5,0.75,0.99,0.9999$ in several double sine-Gordon models.}}
\end{figure}

For completeness we give the particle masses and the kink quantum potential wells:
\[
v^2=1+3a \qquad , \qquad V(x)=\frac{4 (a-1) (3 a+1) \left((15 a+1) \cosh \left(2 \sqrt{3 a+1} x\right)-9 a+1\right)}{\left(-(a-1) \cosh \left(2 \sqrt{3a+1} x\right)+7 a+1\right)^2}  \,\,\, .
\]
We shall compute the one-loop quantum correction to the masses of the kinks  (\ref{mixedsoliton}) for several values of the parameter $a$ within the modified asymptotic approach. These one-loop kink mass shifts are shown in Table 5, where the optimized truncation order is also depicted. Note that, depending on $a$, the asymptotic nature of the series fixes a lower or higher optimum number of terms to be considered in the formula (\ref{numericalcorrection}). All these data are represented graphically in the attached Figure.

\begin{table}[ht]
\hspace{0.4cm}\begin{tabular}{|c|c|c|}
\hline
$a$ & $\Delta \tilde{E}/ \hbar m_d$ & $N_t$ \\ \hline
0.0 & -0.318321 & 20 \\ \hline
0.1 & -0.370625  & 3 \\ \hline
0.2 & -0.429882  & 5 \\ \hline
0.3 & -0.495113  & 5 \\ \hline
0.4 & -0.565305  & 6 \\ \hline
0.5 & -0.637682  & 7 \\ \hline
\end{tabular} \hspace{1cm}
\begin{tabular}{|c|c|c|}
\hline
$a$ & $\Delta \tilde{E}/ \hbar m_d$& $N$ \\ \hline
0.6 & -0.716205  & 8 \\ \hline
0.7     & -0.79149  & 7 \\ \hline
0.8     & -0.90928  & 7 \\ \hline
0.9     & -0.98757  & 10 \\ \hline
0.99    & -1.09687  & 12 \\ \hline
0.999   & -1.11267  & 12 \\ \hline
\end{tabular} \hspace{1cm}
\begin{tabular}{c}
\includegraphics[height=3.5cm]{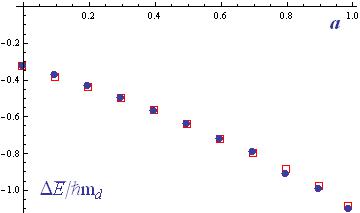}
\end{tabular}
\caption{\textit{One-loop kink mass shifts in the family of double sine-Gordon models for several values of the parameter $a$ (left) and their graphical representation (right).}}
\end{table}

Bearing in mind that at the $a\rightarrow 1$ limit the kink solution (\ref{mixedsoliton}) is formed by two solitons of the re-scaled sine-Gordon model, it is tempting to interpret the quantum correction $\Delta E$ at this limit as the double of the quantum correction of the sine-Gordon soliton. In particular, we recall that $\Delta E_{\rm sG}^{(\pi)} = -0.636894 \hbar m$. We observe behavior similar to this in Table 5, although the accumulation near the point $a=1$ does not allow us to obtain the answer with complete precision.

\section{Summary and outlook}

In a very brief summary and referring to outlooks we wish to stress again that we have proposed a modification of the Gilkey-DeWitt
heat kernel expansion when addressing differential operators with non-trivial algebraic kernels. The new procedure
has been implemented in several one-component scalar field theory models with both computational and conceptual gains
with respect to the application in the same models of the standard procedure. We believe that the translation of these
methods to more complicated models such as $N$-component scalar field theory and/or Abelian Higss will be even more effective.

\section*{Acknowledgements}

We warmly thank our collaborators in previous research into this topic, W. Garcia Fuertes, M. Gonzalez Leon and M. de la Torre Mayado, for illuminating conversations about different aspects of this subject.

We also gratefully acknowledge that this work has been partially financed
by the Spanish Ministerio de Educacion y Ciencia (DGICYT) under grant: FIS2009-10546.

\section*{Appendix}

In this appendix we display a Mathematica code, which automatizes the computation of the quantum correction to the kink mass in one-component scalar field theory by applying formula (\ref{numericalcorrection}), derived from the modified asymptotic series approach. The algorithm is divided into three subroutines: the identification of the density coefficients $c_n(x,x)={}^{0}C_{n}(x)$ by means of (\ref{capitalAcoefficients}); the computation of the Seeley coefficients by integrating the density coefficients, finally evaluating the formula (\ref{numericalcorrection}) to obtain an estimation of the kink mass quantum correction.
\begin{itemize}
\item Calculation of the $c_n(x,x)$ densities.

The following Mathematica code

\vspace{0.2cm}

\hspace{1cm}\begin{minipage}{14cm}
\texttt{\footnotesize
densitycoefficients[potential\_, vacuum1\_, vacuum2\_, nmax\_] :=
Module[\{var1, var2, var3, tomax, d1, v, v0, oper, f0, f6, x7, coef, k, coa, j, co\}, (var1[ph1\_] = potential /. \{y -> ph1\}; var2[ph1\_] = Simplify[PowerExpand[Sqrt[2 var1[ph1]]]]; var3[ph1\_] = Sign[var2[(vacuum1 + vacuum2)/2]] var2[ph1]; coef = \{\}; v[x\_] = Simplify[(D[var1[ph1], \{ph1, 2\}]) /. \{ph1 -> ph1[x]\}]; v0 = Sqrt[Simplify[(D[var1[ph1], \{ph1, 2\}]) /. \{ph1 -> vacuum1\}]]; f0[x\_] = (var3[ph1]/(Sqrt[Integrate[var3[ph1], \{ph1, vacuum1, vacuum2\}]])) /. \{ph1 -> ph1[x]\}; d1[fun\_] :=     Simplify[(D[fun, x]) /. \{ph1'[x] -> var3[ph1[x]]\}]; oper[fu8\_, n1\_] := Simplify[Nest[f6, x7, n1] /. \{f6 -> d1, x7 -> fu8\}]; tomax = 2 nmax; For[k = 0, k < tomax + 0.5, coa[0, k] = 0; k++]; coa[0, 0] = 1; co[n\_, k\_] := Simplify[(1/(n + k)) (coa[n - 1, k + 2] - Sum[Binomial[k, r5] oper[v[x] - v0\^{}2, r5] coa[n - 1, k - r5], \{r5, 0, k\}] - 2 v0 f0[x] oper[f0[x], k] KroneckerDelta[0, n - 1] - f0[x] oper[f0[x], k] (1 + 2 k) (2\^{}n (v0)\^{}(2 n - 1))/((2 n - 1)!!))]; For[j = 1, j < nmax + 0.5, tomax = tomax - 2; For[k = 0, k < tomax + 0.5, coa[j, k] = co[j, k]; If[k == 0, coef = Append[coef, coa[j, 0]]]; 		 k++]; j++]; Return[coef])]; }
\end{minipage}

\vspace{0.2cm}

\noindent defines the module \texttt{densitycoefficients[potential\_, vacuum1\_, vacuum2\_, nmax\_]}, which performs the work of calculating the coefficients $c_n(x,x)$. The arguments of this computational function are \texttt{potential}, the $U(y)$ potential written by prescription as a function of the $y$ variable, \texttt{vacuum1} and $\texttt{vacuum2}$, the two vacua connected by the kink solution in increasing order, and \texttt{nmax}, the $N$-order truncation chosen in the computation of $\Delta E(\phi_K; N)$ .

\item Calculation of the Seeley coefficients.

The Mathematica module \texttt{seeleycoefficients[potential\_, vacuum1\_, vacuum2\_, nmax\_]} depends on the same arguments as the previous one

\vspace{0.2cm}

\hspace{1cm}\begin{minipage}{14cm}
\texttt{\footnotesize
seeleycoefficients[potential\_, vacuum1\_, vacuum2\_, nmax\_] := Module[\{coef, densi, f, f1, f2, a = \{\},j\}, (coef = densitycoefficients[potential, vacuum1, vacuum2, nmax]; f1[y\_] = Simplify[PowerExpand[Sqrt[2 potential]]]; f2[y\_] = Sign[f1[(vacuum1 + vacuum2)/2]] f1[y]; For[j = 1, j < nmax + 0.5, f[y\_] = Simplify[(coef[[j]] /. \{ph1[x] -> y\})/f2[y]]; a = Append[a, Integrate[f[y], \{y, vacuum1, vacuum2\}]]; j++]; Return[a])];
}
\end{minipage}

\vspace{0.2cm}

\noindent and provides us with the value of the Seeley coefficients. This subroutine calls the previous function in order to accomplishes its task.

\vspace{0.2cm}

\item Estimation of the quantum correction.

The subroutine \texttt{quantumcorrection[potential\_, vacuum1\_, vacuum2\_, nmax\_]}

\vspace{0.2cm}

\hspace{1cm}\begin{minipage}{15cm}
\texttt{\footnotesize
quantumcorrection[potential\_, vacuum1\_, vacuum2\_, nmax\_] := Module[\{v0, corr, a\}, (v0 =  Sqrt[Simplify[(D[potential, \{y, 2\}]) /. \{y -> vacuum1\}]]; a = Chop[seeleycoefficients[potential, vacuum1, vacuum2, nmax]]; corr = -(v0/Pi) - (1/(8. Pi)) (Sum[(a[[n]] (v0\^{}(-2 n + 2)) Gamma[n - 1]), \{n, 2, nmax\}]); Return[corr])];
}\end{minipage}

\vspace{0.2cm}

\noindent completes the work by providing us with the one-loop kink mass quantum correction in the (1+1) dimensional scalar field theory model characterized by the potential term $U(y)$. This function calls the two previous ones.
\end{itemize}

The KinkMassQuantumCorrection\_Modified.nb file containing this Mathematica code can be download at the web page http://campus.usal.es/$\sim$mpg/General/Mathematicatools, which includes examples and demos. We recommend this option in order to avoid transcription errors in the code.

\end{document}